\documentclass[prl,twocolumn,showpacs,preprintnumbers,amsmath,amssymb]{revtex4}

\usepackage{graphicx}
\usepackage{dcolumn}
\usepackage{bm}
\usepackage{color}

\newcommand{\mapright}[1]{\smash{\mathop{\hbox to 1.0cm{\rightarrowfill}}\limits^{#1}}}

\begin{document}

\preprint{}

\title{ Josephson $\pi$-state in a ferromagnetic insulator}

\author{Shiro Kawabata,$^{1}$ Yasuhiro Asano,$^{2}$ Yukio Tanaka,$^{3}$ Alexander A. Golubov$^{4}$ and Satoshi Kashiwaya$^{5}$}

\affiliation{
$^1$Nanotechnology Research Institute (NRI), National Institute of Advanced Industrial Science and Technology (AIST), and JST-CREST,  Tsukuba, Ibaraki, 305-8568, Japan
\\
$^2$Department of Applied Physics, Hokkaido University,
Sapporo, 060-8628, Japan
\\
$^3$Department of Applied Physics, Nagoya University,
Nagoya, 464-8603, Japan
\\
$^4$Faculty of Science and Technology and MESA+ Institute of Nanotechnology, University of Twente, 7500 AE, Enshede, The Netherlands
\\
$^5$Nanoelectronics Research Institute (NeRI), AIST, Tsukuba, Ibaraki, 305-8568, Japan
}

\date{\today}

\begin{abstract}
We predict anomalous atomic-scale 0-$\pi$ transitions in a Josephson junction with a 
ferromagnetic-insulator (FI) barrier.
The ground state of such junction alternates between 0- and $\pi$-states when thickness of FI
is increasing by a single atomic layer.
We find that the mechanism of the 0-$\pi$ transition can be attributed to thickness-dependent
phase-shifts between the wave numbers of electrons and holes in FI.
Based on these results, we show that stable $\pi$-state can be realized in junctions
based on high-$T_c$ superconductors with La$_2$BaCuO$_5$ barrier.
\end{abstract}

\maketitle

%
%
%

The developing field of superconducting spintronics
comprises a plenty of fascinating phenomena that may complement
nonsuperconducting spintronics devices~\cite{rf:Zutic}.
Mesoscopic hybrid structures consisting of superconducting and magnetic materials have attracted
considerable attention as devices with novel functionalities~\cite{rf:Buzdin1}.
One of most interesting effects is the formation of $\pi$-states in
superconductor/ferromagnetic-metal/superconductor (S/FM/S) Josephson junctions~\cite{rf:Bulaevskii}.
Under appropriate conditions a ferromagnet can become a $\pi$-phase shifter, providing 
the phase difference $\phi=\pi$ between two superconductors in the ground state
in contrast to $\phi=0$ in ordinary Josephson junctions.
Recently a quiet qubit based on S/FM/S $\pi$-junction~\cite{rf:Ioffe} has been
suggested as a promising device to realize quantum computation because
the spontaneously generated two-level system in this structure is robust against
decoherence due to external fluctuations.
However, low energy quasiparticle excitations in a FM provide strong dissipation~\cite{rf:Schon}.
Therefore Josephson $\pi$ junctions with a nonmetallic interlayers are highly
desired for qubit applications~\cite{rf:Kawabata1}.
Moreover, from the fundamental view point, the Josephson transport through a $ferromagnetic$ $insulator$ (FI) has been studied based on phenomenological models~\cite{rf:SFIS} and not yet been explored explicitly.

In this Letter, we study theoretically the Josephson effect
in superconductor/ferromagnetic-insulator/superconductor (S/FI/S)
junctions using the tight-binding model.
We show that the ground state in such structures alternates between the 0- and $\pi$-states
when the thickness of a FI ($L_F$) is increasing by a single atomic layer.
This remarkable effect originates from the characteristic band
structure of a FI. Quasiparticles in the electron and hole branches
acquire different phase shifts while propagating across a FI.
We will show that the phase difference is exactly $\pi  L_F$ due to the band structure of a FI,
thus providing the atomic-scale 0-$\pi$ transition.
This mechanism is in striking contrast to the proximity induced  $0$-$\pi$ transition in conventional S/FM/S junctions.
On the basis of the obtained results, we predict a stable $\pi$-state in a
Josephson junction based on high-$T_c$ superconductors with a La$_2$BaCuO$_5$ barrier, 
where electric current flows along the $c$ axis
of cuprates. 
%
%
%
%
%
%
\begin{figure}[b]
\begin{center}
\includegraphics[width=8.5cm]{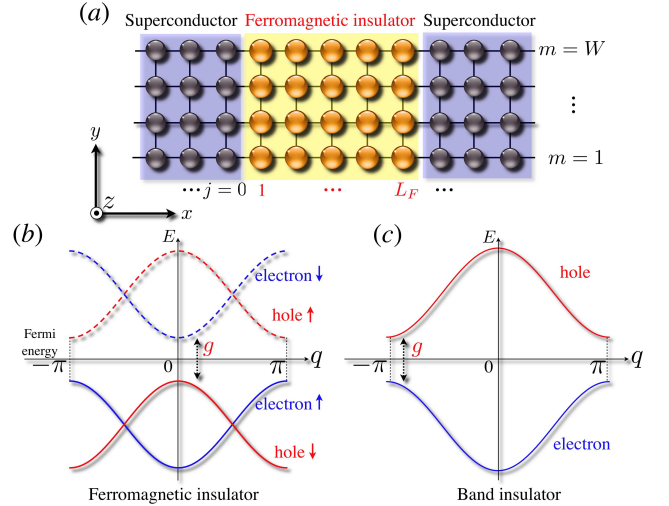}
\end{center}
\caption{(color online) (a)
The Josephson junction with a ferromagnetic-insulator (FI) barrier
on the tight-binding lattice.
The magnetic moment in FI is chosen along the $z$ axis in spin space.
The band structure of a FI (b) and of standard band insulator (c) in the
Bogoliubov-de Gennes picture. The dispersion for a hole with spin $\sigma$ is
obtained as a mirror image of the dispersion for an electron with spin
$\sigma$ with respect to Fermi energy.
 }
\label{fig1}
\end{figure}

Let us first consider an S/FI/S junction in the two-dimensional tight-binding model as shown
in Fig.~1(a). The vector $\boldsymbol{r}=j{\boldsymbol{x}}+m{\boldsymbol{y}}$ points
to a lattice site, where ${\boldsymbol{x}}$ and ${\boldsymbol{y}}$ are unit vectors
in the $x$ and $y$ directions, respectively.
The thickness of a FI layer is $L_F$.
In the $y$ direction, we apply the hard wall boundary condition for the number of
lattice sites being $W$.
Electronic states in a superconductor are described by the
mean-field Hamiltonian,
%
$
{\cal H} =(1/2)
\sum_{\boldsymbol{r},\boldsymbol{r}^{\prime }  \in \text{S}    }%
( \tilde{c}_{\boldsymbol{r}}^{\dagger }\;\hat{h}_{\boldsymbol{r},%
\boldsymbol{r}^{\prime }}\;\tilde{c}_{\boldsymbol{r}^{\prime }}^{{}}-%
\overline{\tilde{c}_{\boldsymbol{r}}}\;\hat{h}_{\boldsymbol{r},\boldsymbol{r}%
^{\prime }}^{\ast }\;\overline{\tilde{c}_{\boldsymbol{r}^{\prime }}^{\dagger
}})
+(1/2)
\sum_{\boldsymbol{r}\in \text{S}} ( \tilde{c}_{%
\boldsymbol{r}}^{\dagger }\;\hat{\Delta}\;\overline{\tilde{c}_{\boldsymbol{r}%
}^{\dagger }}-\overline{\tilde{c}_{\boldsymbol{r}}}\;\hat{\Delta}^{\ast }\;%
\tilde{c}_{\boldsymbol{r}} )
$
with
$\overline{\tilde{c}}_{\boldsymbol{r}}=\left( c_{\boldsymbol{r}%
,\uparrow },c_{\boldsymbol{r},\downarrow }\right) $,
 where
$ c_{\boldsymbol{r} ,\sigma }^{\dagger }$ ($c_{\boldsymbol{r},\sigma }^{{}}$)
 is the creation (annihilation) operator of an electron at $\boldsymbol{r}$ with spin $\sigma
(= \uparrow $ or $\downarrow $ ), $\overline{\tilde{c}}$ means the
transpose of $\tilde{c}$,  and $\hat{\sigma}_{0}$ is $2\times 2$ unit matrix.
We introduce the hopping integral $t$ among nearest neighbor sites
and measure the length in the units of the lattice constant $a$.
In superconductors, the Hamiltonian leads
$
\hat{h}_{\boldsymbol{r},\boldsymbol{r}'}= [ -t \delta _{|\boldsymbol{r}-\boldsymbol{r}'|,1}
+
(-\mu_s+4t)\delta_{\boldsymbol{r},\boldsymbol{r}'} ] \hat{\sigma}_{0}
$,
the chemical potential $\mu_s$ is measured from the band bottom
and $\hat{\Delta}=i\Delta \hat{\sigma}_{2}$, where $\Delta$ is the amplitude
of the pair potential in the $s$-wave symmetry and $\hat{\sigma}_{j}$ for $j=1-3$
are Pauli matrices. We describe a FI by
$
\hat{h}_{\boldsymbol{r},\boldsymbol{r}'}= -t \delta_{|\boldsymbol{r}-\boldsymbol{r}'|,1}
\hat{\sigma}_{0}
- (g/2+4t) \delta_{\boldsymbol{r},\boldsymbol{r}'}\hat{\sigma}_3
$,
where $g$ corresponds to a gap of a FI as shown in Fig.~1(b).

The Hamiltonian is diagonalized by the Bogoliubov transformation.
The Andreev bound state consists of subgap states whose
wave functions decay far from the junction interface.
In what follows, we focus on the subspace for
spin-$\uparrow$ electron and spin-$\downarrow$ hole
[the dispersions shown by solid curves in Fig.~1(b)].
In superconductors, the wave function of a bound state is given by
\begin{align}
\Psi_L(\boldsymbol{r})=&\Phi_L \left[
 \left(\begin{array}{c} u \\ v \end{array} \right)
 A e^{-ikj} +
 \left(\begin{array}{c} v \\ u \end{array} \right)B e^{ik^*j}\right] \chi_l(m),\\
\Psi_R(\boldsymbol{r})=&\Phi_R \left[
 \left(\begin{array}{c} u \\ v \end{array} \right)
 C e^{ikj} +
 \left(\begin{array}{c} v \\ u \end{array} \right)D e^{-ik^*j}\right] \chi_l(m),
\end{align}
where $\nu=L$ ($R$) indicates a superconductor in the left (right) hand side,
$\phi_\nu$ is the phase of a superconductor,
$\Phi_\nu=\mathrm{diag} \left(  e^{i\phi_\nu/2} , e^{-i\phi_\nu/2} \right)$,
$u(v)=[ ( 1+(-) \sqrt{E^2-\Delta^2}/E )/ 2 ]^{1/2}$, and $A, B, C$ and $D$
are amplitudes of the wave function for an outgoing quasiparticle.
The wave function in the $y$ direction is $\chi_l(m)=\sqrt{2/W}\sin[l m\pi /(W+1)]$,
where $l$ indicates a transport channel.
The energy $E$ is measured from the Fermi energy
and $k= \cos^{-1} [2-\mu_s/2t  - \cos \{ l \pi / (W+1) \} + i \sqrt{\Delta^2 - E^2} /E]$ is the complex wave number.
These wave functions decay as $e^{-(j-L_F)/\xi_0}$ for $j > L_F$
and $e^{j/\xi_0}$ for $j<0$ with $\xi_0$ being the coherence length.
In a FI, the wave function is given by
\begin{align}
\Psi_{FI}(\boldsymbol{r})=&
\left[ \left(\begin{array}{c} f_L e^{-iq_ej} \\ g_L e^{-iq_hj}\end{array} \right)
+\left(\begin{array}{c} f_R e^{iq_ej} \\ g_R e^{iq_hj}\end{array} \right)\right] \chi_l(m),\\
 q_e=&\pi + i\beta_\uparrow, \label{wn1}\\
   q_h=&0+i\beta_\downarrow,\label{wn2}
\end{align}
where $\cosh\beta_\uparrow= 1 + E/2t+g/4t + \cos [ l \pi / (W+1) ]- \cos [ W \pi / (W+1) ]$,
 $\cosh \beta_\downarrow= 1 + E/2t+g/4t - \cos [ l \pi / (W+1) ]- \cos [  \pi / (W+1) ]$, and
$f_L, f_R, g_L$ and $g_R$ are amplitudes of wave function in a FI.
The Andreev levels $\varepsilon_{n,l} (\phi=\phi_L-\phi_R)$ [$n=1, \cdots, 4$]
can be calculated from boundary conditions
$\Psi_L(\lambda,m)=\Psi_{FI}(\lambda,m)$ and
$\Psi_R(L_F+\lambda,m)=\Psi_{FI}(L_F+\lambda,m)$ for $\lambda =0$ and 1.
The Josephson current is related to $\varepsilon_{n,l}$ via
$ I_J (\phi) =(2e / \hbar) \sum_{n,l}  \left[ \partial \varepsilon_{n,l}
(\phi)/ \partial \phi \right]  f  \left( \varepsilon_{n,l} (\phi) \right),
$
where $ f\left( \varepsilon  \right)$ is the Fermi-Dirac distribution function.
The Josephson critical current  $I_C$ is defined by $I_C = I_J(\pi/2)$.

\begin{figure}[b]
\begin{center}
\includegraphics[width=8.7cm]{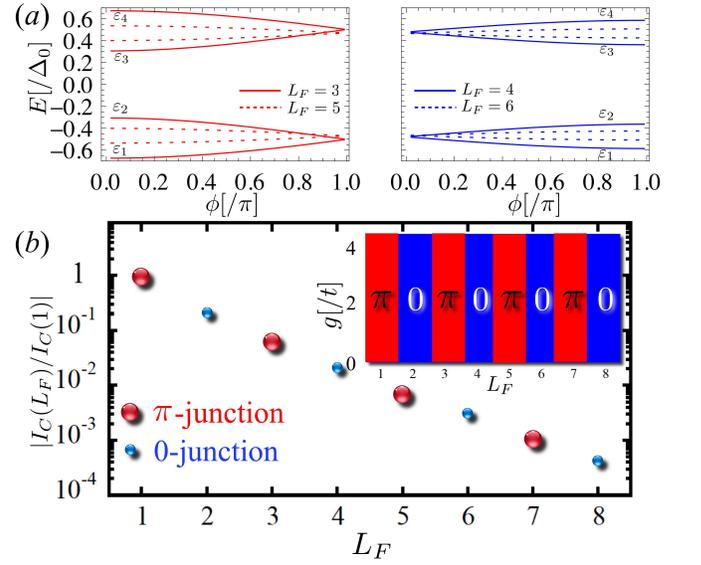}
\end{center}
\caption{(color online)
(a) The Andreev levels $\varepsilon_{i} \equiv \varepsilon_{i,1}$ are plotted as functions of $\phi$ for an one-dimensional S/FI/S junction with $W=1$.
 Left  (right)  panel shows the results for odd $L_F$ (even $L_F$) with
$g=0.5t$.
(b) Josephson critical current $I_C$ at $T=0.01 T_c$ as a function of the thickness $L_F$ for
$W=1$ and $g=0.5 t$. The large red (small blue) circles indicate the $\pi$(0) junction.
In the inset, the 0-$\pi$ phase diagram on the $g$-$L_F$ plane is shown for
a two-dimensional  S/FI/S junction with $W=10$ at $T=0.01 T_c$, where the 
red and blue regimes correspond to the $\pi$- and 0-states, respectively.}
\label{fig2}
\end{figure}

In Fig.~\ref{fig2}(a), we first show the Andreev levels $\varepsilon_{n,1} \equiv  \varepsilon_{n}$ for odd $L_F$ ($=3$ and 5) and even  $L_F$ ($=4$ and 6) with $W=1$,
$\mu_s=2t$, and $\Delta=0.01t$.
The results show that the ground state for odd $L_F$ is at $\phi=\pi$, whereas that for even $L_F$ is at $\phi=0$.
This atomic-scale 0-$\pi$ transition persists  even if we increase $L_F$ and $W$.
In Fig.~2(b), we show the Josephson critical current as a function of $L_F$ for $W=1$.
Temperature $T$ is set to be $0.01T_c  \ll T_c$, where $T_c$ is the transition temperature of a superconductor.
The $\pi$(0)-state is always more stable than the 0($\pi$)-state when the thickness of FI is an odd(even) integer.
The reason is as follows.
At low temperatures, only the Andreev levels below the Fermi energy i.e.,
$\varepsilon_{1}$ and
$\varepsilon_{2}$, contribute to $I_C$ [see Fig. 2(a)].
In the odd (even) $L_F$ cases, the $\pi$- (0-) state is stable because of
$  \partial_\phi \varepsilon_1   |_{\phi=\pi/2}> (<) 0 $,
$ \partial_\phi \varepsilon_2   |_{\phi=\pi/2}< (>) 0 $, and
$ |  \partial_\phi \varepsilon_2   |_{\phi=\pi/2} |>  |\partial_\phi \varepsilon_1   |_{\phi=\pi/2} |$.
The atomic-scale 0-$\pi$ transition is  insensitive to $W$ and material parameters such as $\mu_s$, $g$, and $\Delta$.
As an example, inset of Fig.~2(b) shows the phase diagram on the $g$-$L_F$ plane for $W=10$.

The mechanism of the 0-$\pi$ transition in a FI is very different from that
in a FM. The key feature is expressed by the wave number of a quasiparticle
in a FI as shown in Eqs.~(\ref{wn1}) and (\ref{wn2}),
where $q_e$ and $q_h$ are the wave numbers for an electron spin-$\uparrow$
and a hole spin-$\downarrow$, respectively.
The real parts of $q_e$ and $q_h$ reflect the wave number at the $q$ points, where energy
is closest to the Fermi energy, and differ by $\pi$ from each other .
 As shown in Fig. 1(b), the real part of $q_e$ is
$\pi$ because the top of the electron band is located at $q=\pi$.
On the other hand, the real part of $q_h$  is 0 because the top of the hole
band is at $q=0$. This is the origin of the difference between $q_e$ and $q_h$ which
accounts the atomic-scale 0-$\pi$ transition.
When we consider a usual band insulator as shown in Fig. 1(c),
we always obtain $q_e=q_h$ and their real parts equal $\pi$
because both the top of the electron band and the bottom of the hole band are located at $q=\pi$.
As a consequence, 0-state is always stable in usual band insulators.
Thus we conclude that the characteristic band structure of a FI is the origin
of atomic-scale 0-$\pi$ transitions.
These features basically remain unchanged
even when we consider Josephson junctions in higher dimensions. In such junctions, however, the appearance of 
0-$\pi$ transitions depends on relative directions between the
current and the crystalline axis. We will address this issue below.
It should be emphasized that peculiar results presented above cannot be described by the standard quasiclassical Green's function method~\cite{rf:Fogelstrom} where band structure structure far from the Fermi energy is is ignored.

Let us reconsider atomic-scale 0-$\pi$ transitions from a different view point of quasiparticle
transmission coefficient.
In the high barrier limit ($g \gg t$), the Josephson critical current is perturbatively given by
$I_C \propto T_\downarrow^* T_\uparrow$~\cite{rf:Kawabata1}.
Here $T_{\uparrow(\downarrow)}$ is a transmission coefficient of a FI for
an electron with spin-$\uparrow$ (-$\downarrow$).
By using the transfer-matrix method~\cite{rf:Datta}, $T_\sigma$ for
one-dimensional junctions can be obtained analytically
$
  T_\sigma \approx
\alpha_{L_F}\left(  \rho_\sigma t  /  g \right)^{L_F}.
$
Here $\rho_{\uparrow(\downarrow)} = -(+)1$ and $\alpha_{L_F}$ is a spin-independent complex constant.
We immediately find $T_\uparrow/T_\downarrow = (-1)^{L_F}$.
Thus the relative phase
of $T_\sigma$ between spin-$\uparrow$ and spin-$\downarrow$
 is $\pi$ (0) for the odd (even) number of $L_F$.
As a consequence, the sign of $I_C \propto  (-1)^{L_F} $ becomes negative for odd $L_F$ and positive for even $L_F$.
In other words, a FI acts as a $\pi$-phase-shifter for the spin-$\uparrow$ electron for odd $L_F$.

The transfer-matrix method in real space also enables us to extend the calculations to
another magnetic materials.
Up to now, we have considered uniform magnetic moment in FI, which can be schematically expressed by
S/$\uparrow_1 \uparrow_2 \cdots \uparrow_{L_F}$/S or S/$\downarrow_1 \downarrow_2
\cdots \downarrow_{L_F}$/S. The arrows $\uparrow_j$ and
$\downarrow_j$ indicate the $z$-axis magnetization at $j$.
We can extend the above simple analysis to the arbitrary
magnetization configuration, $e.g.,$ a
random alignment described by S/$\downarrow_1 \uparrow_2 \downarrow_ 3 \cdots
\uparrow_{L_F-2} \uparrow_{L_F-1} \downarrow_{L_F}$/S.
In such junctions, we find
$I_C \sim \prod_{i=1,L_F} T_{i, \mathrm{\downarrow}}^* T_{i, \mathrm{\uparrow}}
 =  \prod_{i=1,L_F} (-1)=(-1)^{L_F}$,
where $T_{i,\sigma}$ is the transmission coefficient of an FI layer at $i$.
Therefore we obtain a noticeable result, $i.e.$,
the sign of
 $I_C$
is independent of magnetization configurations and is negative (positive) for odd (even) $L_F$.
The appearance of the atomic-scale 0-$\pi$ transition has been also predicted in
S/antiferromagnetic-interlayer/S junctions~\cite{rf:Barash}.
In their theory, however, the antiferromagnetic configuration
is found to be essential for the atomic-scale transition.
On the other hand, we conclude that the magnetization symmetry is not necessary and that
the $\pi$-phase difference between $T_\uparrow$ and $T_\downarrow$ is an essential feature
for the atomic-scale transition.
Therefore our analysis provides more general view for the physics of the atomic scale 0-$\pi$ transition.

Finally, we show the possibility of the atomic-scale 0-$\pi$ transition in a three-dimensional
junction using realistic materials.
Here we focus on La$_2$BaCuO$_5$ (LBCO)~\cite{rf:Mizuno} which is an important representative FI in spintronics.
According to a first-principle band calculation~\cite{rf:LBCO}, the bottom of the minority spin band is at the $\Gamma$ point
whose wave number is $(k_a, k_b, k_c) = (0,0,0)$, where $k_j$ for $j=a,b,$ and $c$ is the
wave number along $j$ axis (see Fig.~6 in Ref.~\onlinecite{rf:LBCO}).
The mirror image of the minority spin band with respect to the Fermi energy corresponds to
the hole band with minority spin in the Bogoliubov-de Gennes picture.
Thus the top of the minority spin hole band is at the $\Gamma$ point.
On the other hand, top of the majority spin band is at the $Z$ point with
$(k_a, k_b, k_c) = (0,0,\pi)$.
Thus we can predict that the $\pi$ state would be possible if one fabricates a Josephson
junction along $c$ axis as shown in Fig.~4(a). Note that it is impossible to
realize the $\pi$-state if current flows in the $ab$-plane.
This is because wave numbers in $ab$-plane at the bottom of the minority spin band and
those at the top of the majority spin band are given by the same wave number
$(k_a, k_b) = (0,0)$~\cite{rf:LBCO}.

From the perspectives of the S/FI interface matching and the high-temperature device-operation,
the usage of high-$T_c$ cuprate superconductors (HTSC), e.g.,
YBa${}_2$Cu${}_3$O${}_{7-\delta}$ and La${}_{2-x}$Sr${}_x$CuO${}_4$ (LSCO) is desirable.
Recent development of the pulsed laser deposition technique enables layer-by-layer
epitaxial-growth of oxide superlattices~\cite{rf:Prellier}.
In order to show the possibility of $\pi$-coupling in such realistic HTSC junctions, we have calculated the $c$-axis Josephson critical current $I_C$ based on a three-dimensional tight-binding model with $L_a$ and $L_b$ being the numbers of lattice sites in $a$ and $b$ directions [Fig. 3 (a)].
In the calculation we have taken into account the $d$-wave order-parameter symmetry in HTSC, i.e., $\Delta= \Delta_d (\cos k_x a- \cos k_y a) / 2$. The tight-binding parameters $t$ and $g$  have been determined  by fitting to the first-principle band structure calculations along the line from $\Gamma$ to $Z$ point~\cite{rf:LBCO}.
Figure 3 (c) shows the thickness $L_F$ dependence of $I_C$ at $T=0.01 T_c$ for a LSCO/LBCO/LSCO junction with $g/t=20$, $\Delta_d/t=0.6$, and $L_a=L_b=100$.
As expected, the atomic-scale  0-$\pi$ transitions can be realized in such oxide-based $c$-axis stack junctions.
\begin{figure}[tb]
\begin{center}
\includegraphics[width=8.8cm]{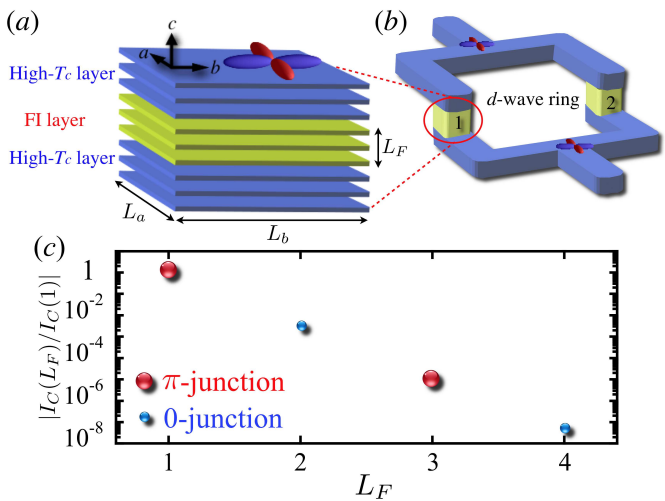}
\end{center}
\caption{
(color online) (a) Schematic picture of a $c$-axis stack high-$T_c$
superconductor/FI/high-$T_c$ superconductor Josephson junction and (b) a $d$-wave ring to detect the $\pi$-junction behavior experimentally.
(c) The Josephson critical current $I_C$ as a function of the FI thickness $L_F$ at $T=0.01 T_c$ for a $c$-axis stack LSCO/LBCO/LSCO junction with $g/t=20$, $\Delta_d/t=0.6$, and $L_a=L_b=100$.
The large red (small blue) circles indicate the $\pi$(0) junction.}
\label{fig3}
\end{figure}

The experimental detection of the $\pi$-junction is possible by using a superconducting ring
which contains two Josephson junctions as shown in Fig.~\ref{fig3}(b).
When both junctions are in 0- (or $\pi$-) state at the same time,
the critical current of the ring reaches its maximum at zero
external magnetic flux. On the other hand, the critical current reaches its minimum
at zero magnetic flux when the 0 state is stable in one junction and $\pi$ is stable
in the other~\cite{rf:Sigrist}.

From the  view point of qubit applications, it is important to note that the harmful influence  of  midgap Andreev resonant states~\cite{rf:Tanaka,rf:KawabataMQT1} and nodalquasiparticles due to the $d$-wave symmetry on the macroscopic quantum dynamics  in $c$-axis HTSC junctions~\cite{rf:Kleiner} is found to be weak, both theoretically~\cite{rf:KawabataMQT2} and experimentally~\cite{rf:MQTexperiment}.
Therefore we conclude that  HTSC/LBCO/HTSC
$\pi$-junctions would be promising candidates as basic-elements for  quiet qubits.

In summary, we have studied the Josephson effect in S/FI/S junctions based on the tight-binding model.
We predict the formation of the atomic-scale 0-$\pi$ transitions in such junctions.
This result is insensitive to the material parameters such as the gap $g$ of the FI and the superconducting gap $\Delta$, indicating that it is a robust and universal phenomenon.
Our findings suggest the way of realizing  {\it ideal quiet qubits} which possess both the quietness and the weak quasiparticle-dissipation nature.

We  would like to thank J. Arts, A. Brinkman, M. Fogelstr\"om, T. Kato, P. J. Kelly, T. L\"ofwander, T. Nagahama, F. Nori, J. Pfeiffer, A. S. Vasenko, and M. Weides for useful discussions.
This work was  supported by CREST-JST, a Grant-in-Aid for Scientific Research from the Ministry of Education, Science, Sports and Culture of Japan (Grant No. 19710085),  and  NanoNed (Grant TCS.7029).

\end{document}